\documentclass[letterpaper]{article} % DO NOT CHANGE THIS
\usepackage{aaai2026}  % DO NOT CHANGE THIS
\usepackage{times}  % DO NOT CHANGE THIS
\usepackage{helvet}  % DO NOT CHANGE THIS
\usepackage{courier}  % DO NOT CHANGE THIS
\usepackage[hyphens]{url}  % DO NOT CHANGE THIS
\usepackage{graphicx} % DO NOT CHANGE THIS
\usepackage{natbib}  % DO NOT CHANGE THIS AND DO NOT ADD ANY OPTIONS TO IT
\usepackage{caption} % DO NOT CHANGE THIS AND DO NOT ADD ANY OPTIONS TO IT
\usepackage{algorithm}
\usepackage{algorithmic}

\usepackage{newfloat}
\usepackage{listings}
\DeclareCaptionStyle{ruled}{labelfont=normalfont,labelsep=colon,strut=off} % DO NOT CHANGE THIS
\floatstyle{ruled}
\newfloat{listing}{tb}{lst}{}
\floatname{listing}{Listing}
\title{Conversational AI for Social Good (CAI4SG): An Overview of Emerging Trends, Applications, and Challenges}
\author {
    Yi-Chieh Lee\textsuperscript{\rm 1},
    Junti Zhang\textsuperscript{\rm 2},
    Tianqi Song\textsuperscript{\rm 1},
    Yugin Tan\textsuperscript{\rm 1}
}
\affiliations {
    \textsuperscript{\rm 1}Computer Science, School of Computing, National University of Singapore\\
    \textsuperscript{\rm 2} Institute of Data Science, National University of Singapore\\
    yclee@nus.edu.sg, juntizhang@u.nus.edu, tianqi\_song@u.nus.edu, tanyugin@nus.edu.sg
}

\begin{document}

\maketitle

\begin{abstract}
The integration of Conversational Agents (CAs) into daily life offers opportunities to tackle global challenges, leading to the emergence of Conversational AI for Social Good (CAI4SG). This paper examines the advancements of CAI4SG using a role-based framework that categorizes systems according to their AI autonomy and emotional engagement. This framework emphasizes the importance of considering the role of CAs in social good contexts, such as serving as empathetic supporters in mental health or functioning as assistants for accessibility. Additionally, exploring the deployment of CAs in various roles raises unique challenges, including algorithmic bias, data privacy, and potential socio-technical harms. These issues can differ based on the CA's role and level of engagement. This paper provides an overview of the current landscape, offering a role-based understanding that can guide future research and design aimed at the equitable, ethical, and effective development of CAI4SG.
\end{abstract}

% Uncomment the following to link to your code, datasets, an extended version or similar.
% You must keep this block between (not within) the abstract and the main body of the paper.
% \begin{links}
%     \link{Code}{https://aaai.org/example/code}
%     \link{Datasets}{https://aaai.org/example/datasets}
%     \link{Extended version}{https://aaai.org/example/extended-version}
% \end{links}

\section{Introduction}

Conversational AI (CAI) systems, defined as technologies enabling human-computer interaction through natural language, have evolved from early rule-based dialogue systems to sophisticated contemporary platforms spanning task-oriented applications to open-domain chatbots~\cite{raux2005let,zhou2020design}.
These systems offer distinct advantages including scalability, continuous availability, and cost-effectiveness compared to human operators, while enabling enhanced personalization and privacy preservation~\cite{budzianowski2018multiwoz}.
The global chatbot market has experienced substantial growth, positioning CAI as a transformative technology for diverse applications from customer service to healthcare support.

AI for Social Good (AI4SG) represents the application of AI technologies to address pressing social, environmental, and economic challenges facing humanity~\cite{shi2020artificial}. 
While the field lacks a universally accepted definition, AI4SG is characterized by its focus on generating positive societal outcomes through technological innovation~\cite{berendt2019ai}.
It encompasses diverse application domains that tackle societal challenges, including those previously underexplored by the AI community. 
This inclusive approach aligns with the United Nations Sustainable Development Goals (SDGs)~\cite{biermann2017global} and prioritizes principles of equity, inclusion, and fairness in both research development and deployment.

The intersection of CAI and social good creates unprecedented opportunities for addressing societal challenges through scalable technology. 
CAI systems' natural language interfaces eliminate technological barriers, enabling deployment across diverse populations regardless of technical literacy~\cite{zhou2020design}.
% These platforms demonstrate unique capacity for continuous operation and personalized interaction delivery, making them particularly valuable for underserved communities with limited access to traditional services.
% CAI4SG applications span critical domains including mental health support, educational assistance, disaster response, and accessibility enhancement, positioning CAI as a transformative tool for democratizing essential services and bridging socioeconomic disparities.
CAI for social good (CAI4SG) encompasses chatbots and virtual assistants deployed to advance public well-being, expand access to trustworthy information, and support underserved groups across mental health, public health, education, and humanitarian response. 
The scope thus spans both transformative applications and the safeguards needed to prevent harm while amplifying social impact.

The deployment of CAI in social good contexts introduces unique considerations regarding how these systems interact with diverse communities and deliver essential services.
Inspired by a role-based framework \cite{zhang2025computational}, we analyze CAI4SG applications using two key dimensions: \textbf{AI autonomy} and \textbf{AI emotional engagement}. This approach prioritizes real-world impact over technical specifications, allowing us to better anticipate a system's benefits, risks, and governance needs by examining how it acts on and affects users.

AI autonomy represents the degree of independent decision-making authority delegated to the system. Low-autonomy tools play a more supportive role, providing information or facilitating routine services that streamline user experiences, while high-autonomy systems actively engage the user in collaborative decision-making or problem solving. 
% that can proactively manage interactions and generate contextually appropriate responses without direct oversight~\cite{kuleindiren2022optimizing,lin2022deep}. 
AI emotional engagement reflects the extent to which systems are designed to recognize and respond to users' emotional states through empathic understanding and emotionally resonant communication~\cite{lawrence2024opportunities}.
While low-engagement systems focus on information exchange and task completion, high-engagement systems demonstrate emotional intelligence through validation and empathic responses that can establish meaningful relationships with users. 
These dimensions are particularly significant in social good applications, as they directly influence both the potential benefits, like enhanced accessibility, and associated risks, including accountability concerns and emotional dependency.

This role-based framework delineates the diverse functions and impacts of CAI across social good applications, encompassing a spectrum from functional assistants that perform basic task-oriented services to virtual companions that provide sustained emotional support and relationship-building. 
Importantly, CAI systems may dynamically blend or transition between these roles depending on user needs and contextual requirements. 
This paper aims to provide a concise, high-level synthesis of emerging developments in CAI4SG, highlighting key applications and works while offering a critical synthesis of the current state of the art. 
Through this role-based analytical lens, we suggest future research directions and implications for the broader AI community to develop responsible AI systems.

\section{The Diverse Roles and Transformative Applications of CAI4SG}

By mapping CAIs' functional roles along the axes of autonomy and emotional engagement, we explore key developments and seminal works in diverse societal contexts.
These roles represent points along a dynamic spectrum rather than discrete categories, as CAI systems may transition between functions depending on specific social contexts.
The subsequent sections examine each role's manifestations and implementation opportunities within diverse social good domains.

\subsection{Low Autonomy, Low Emotional Engagement}

When CAIs operate with minimal emotional engagement and low autonomy, they primarily serve instrumental functions that enhance operational efficiency and service delivery across public and social sectors.
These tools provide consistent, scalable accessibility infrastructure that reduces technological barriers and democratizes access to digital services for users with diverse accessibility needs.

\subsubsection{Streamlining Public Sector Services.}
Low-autonomy CAI systems enhance public sector efficiency and accessibility , reducing administrative overheads through automation.
In government services, these systems guide citizens through predefined workflows for benefits applications or permit requests, collecting information within established protocols but requiring human administrators to review critical decisions~\cite{nirala2022survey}. 
Citizen service chatbots handle routine inquiries while escalating complex cases to human operators, and intelligent call routing systems suggest departmental transfers subject to human supervisor approval.
Beyond domestic government workflows, AI systems are also increasingly deployed to support international development governance, for example, by assisting agencies in classifying and monitoring aid contributions toward the Sustainable Development Goals~\cite{park2025classifying}.

\subsubsection{Basic Accessibility Support.}
More broadly, this class of CAIs also enhances accessibility and usability in multiple contexts. 
These systems can execute speech-to-text conversion for hearing-impaired users~\cite{lee2016dialogue} and text-to-speech applications for visually impaired individuals~\cite{isewon2014design}, enhance multilingual communication~\cite{qin2025ai}, and synthesize information retrieval responses ~\cite{yang2025understanding}. 
These applications can independently adapt to different speech patterns, accents, and environmental conditions while maintaining reasonable speed and accuracy in real-time processing.

% Broader implication
\subsubsection{}
Taken together, these applications illustrate that low-autonomy, low-engagement CAIs already exercise transformative leverage through infrastructural roles: 
They expand the institutional surface area at which basic digital access becomes equitably deliverable at the population scale. 
Their value, therefore, lies not only in efficiency gains, but in establishing the baseline socio-technical ``plumbing'' upon which higher-layer social good interventions can be reliably built. 
As such, this quadrant is less about sophisticated interactional intelligence and more about creating dependable, standardized, and interpretable service primitives that future, more autonomous and emotionally engaged deployments can compositionally assemble, inherit, or extend.

% \subsubsection{Multi-Agent Conversational Systems.}

\subsection{Low Autonomy, High Emotional Engagement}

% Combatting Stigma: CAIs can collect qualitative data on sensitive topics like depression stigma, and AI-assisted coding can deconstruct stigma and inform tailored micro-interventions. 
% Emotional Monitoring for Therapists: CAI provides continuous emotional data streams outside of clinical settings, extending the reach of mental health care.
% Fostering Prosociality and Human Well-being: Research indicates that humans helping AI agents can enhance their own well-being, reducing loneliness and increasing positive affect when needs for competence and autonomy are met. This concept of "Reciprocal AI" involves designing AI to genuinely ask for help from humans, leveraging a moderate level of emotional engagement to improve human well-being.

When CAIs operate with limited autonomy yet high emotional engagement, they primarily serve as empathetic listeners that create safe, non-directive spaces for user expression. 
These systems focus on validating emotions and fostering connection rather than generating autonomous decisions, making them particularly valuable in contexts of stigma, vulnerability, and loneliness. 

\subsubsection{Understanding Stigmatized Societal Issues.}
Supportive listener CAIs can collect qualitative narratives on sensitive or taboo issues such as depression or anxiety, enabling users to share experiences without fear of judgment~\cite{lee2023exploring, cui2024exploring, meng2025stigma, song2025interaction}. 
Yet, recent benchmarks demonstrate that generative language models can themselves amplify stigma against marginalized groups~\cite{nagireddy2024socialstigmaqa}, underscoring the need to study supportive listener systems not only as data collectors but also as potential mediators of harm.
With AI-assisted coding, these narratives can be systematically analyzed to deconstruct stigmatizing patterns and inform tailored micro-interventions~\cite{meng2025deconstructing}. In this way, CAIs act as low-barrier entry points for populations hesitant to engage with traditional support services.

\subsubsection{Extending Emotional Care.}
By continuously tracking emotional signals outside formal clinical encounters, supportive listener systems can complement therapists’ work and extend the reach of mental health care~\cite{li2023systematic}. Rather than replacing professionals, they provide contextualized emotional data streams that inform ongoing treatment while maintaining user trust through empathic engagement. Beyond clinical contexts, research also suggests that reciprocal dynamics, where CAIs occasionally request help from users, may enhance prosociality and human well-being~\cite{li2025exploring, zhu2025benefits, chen2025humans}, reducing loneliness and fostering positive affect when users’ needs for competence and autonomy are fulfilled.

% Broader implication
\subsubsection{}
Viewed through a broader lens, this quadrant underscores that the societal contribution of low‐autonomy yet high-engagement CAIs is not decision quality per se, but the affective bandwidth they unlock. 
By reliably scaffolding disclosure, reflection, and emotionally safe self-presentation among users who may otherwise disengage from formal institutions, these systems expand the emotional data surface that mental health and social service ecosystems can meaningfully act upon. 
In this sense, their emerging role is building population-scale emotional observability that can enable downstream, more targeted, and professionally enacted interventions.

\subsection{High Autonomy, Low Emotional Engagement}

When conversational AI systems operate with high autonomy but minimal emotional engagement, they function as efficient information processors and service providers, independently managing complex tasks while maintaining a focus on factual accuracy and systematic response delivery.

\subsubsection{Digital Health Assistants.}
In digital health contexts, CAIs can provide immediate, autonomous pre-diagnostic guidance and health information without requiring direct medical supervision for routine inquiries. 
These systems can independently assess symptom descriptions, recommend appropriate care levels, and direct patients to suitable healthcare services based on established medical protocols~\cite{zhang2025mining}. 
% Operating with minimal emotional engagement, they focus on efficient triage and resource allocation rather than empathetic interaction. 
Such applications significantly reduce burden on healthcare systems by autonomously handling routine health questions, yet the broader autonomous agent literature suggests that such LLM-based orchestrators still lack robustness and reliability for mission-critical deployments~\cite{muthusamy2023towards}, highlighting the need for domain-specific guardrails.

\subsubsection{Combating Misinformation.}
In addressing misinformation spread across digital platforms, CAI systems leverage natural language interaction to effectively engage users who may be exposed to misleading content. 
Through conversational interfaces, these systems enable users to voluntarily submit questionable content for verification, creating a channel for fact-checking and engage in follow-up dialogue to address doubts~\cite{lim2023fact}.
A notable example is the World Health Organization COVID-19 chatbot~\cite{miner2020chatbots}, which engaged millions in interactive dialogue during the pandemic, answering health queries while automatically escalating complex misinformation patterns to human experts. 
The conversational format proves particularly effective because it allows for real-time clarification of misunderstandings, enables users to pose follow-up questions, and facilitates the delivery of authoritative information in familiar, accessible communication patterns that mirror human-to-human information sharing.

\subsubsection{}
More broadly, this quadrant highlights a distinct pathway for CAI4SG impact: high-autonomy, low-engagement systems function as service utilities that execute continuous, protocol-aligned control over information flows at population scale. 
Their primary contribution is not interactional depth but system-level throughput and providing consistent triage, enforcing evidence-based guidance, and attenuating misinformation propagation without requiring human oversight for routine cases. 
In this sense, the emerging trend in this quadrant is the operationalization of CAI as autonomous socio-technical infrastructure that stabilizes public information environments and expands access to core services.

\subsection{High Emotional Engagement, High Autonomy}

When CAI systems operate with both high emotional engagement and significant autonomy, they function as social partners capable of establishing meaningful relationships while independently managing complex, long-term interactions.

\subsubsection{Mental Health Companions.}
Research indicates that CAI-mediated mental health interventions can significantly reduce symptoms of depression and anxiety while providing round-the-clock availability for individuals in crisis~\cite{fitzpatrick2017delivering}.
The stigma-free nature of AI interactions enables users to engage in self-disclosure without fear of judgment, particularly benefiting populations who might otherwise avoid seeking mental health services due to cultural barriers or social stigma~\cite{zhang2025dark}.
Moreover, CAIs also demonstrate particular value through their capacity to identify subtle patterns in mood, behavior, and cognitive states that episodic clinical encounters might miss~\cite{rathnayaka2022mental}.
This highlights the significant potential of CAI in establishing meaningful long-term therapeutic relationships~\cite{zhang2019clinically}.

\subsubsection{Personalized Learning.}
CAI systems with high emotional engagement and autonomy revolutionize educational delivery through real-time adaptation to individual learning styles, cognitive abilities, and emotional states.
Research demonstrates significant potential for AI tutoring systems to address educational inequities across diverse populations, from students with special educational needs~\cite{zawacki2019systematic} to older adult learners seeking AI literacy~\cite{tang2025ai}.
Personalized CAI can address these needs by providing patient, adaptive instruction that accommodates different learning paces and cognitive abilities, offering immediate clarification and practical examples tailored to individual contexts.

\subsubsection{Social Support Systems.}
Beyond individual therapy, high-autonomy CAIs can strengthen users’ connectedness by linking them to peer networks, support groups, and community resources~\cite{geng2025beyond, bae2021social}.
For example, CAIs can help older adults sustain social participation by recommending age-friendly events, connecting them with peers who share similar experiences, or mediating intergenerational communication~\cite{du2025ai}.
In marginalized contexts, CAIs can lower barriers to resources and amplify users’ voices~\cite{lee2025amplifying}.
Through these functions, they move from personal aides to enablers of collective resilience.

\subsubsection{}
Taken at scale, this quadrant suggests a qualitatively different mode of CAI4SG impact: systems become relational intervention engines that can simultaneously sustain emotionally meaningful engagement and autonomously enact long-horizon support trajectories. 
The broader implication is that the locus of intelligence shifts from single-shot assistance toward the continual shaping of users’ affective, cognitive, and social states over time. 
As this class of systems matures, the central research problem is no longer whether they can help, but how to govern the autonomy of long-term, emotionally entangled interventions such that benefits are realized without producing new forms of dependency, manipulation, or opaque influence.

\section{Critical Synthesis: Challenges and Ethical Considerations Across CAI Roles}

This section will examine how challenges and ethical considerations manifest across different CAI4SG roles, and proposes potential solutions to address them.

\subsection{Emotional, Ethical, and Social Risks}

Our role-based analysis reveals that CAI roles with high emotional engagement present the most significant ethical and social risks, as their capacity for emotional manipulation and relationship simulation creates unique vulnerabilities among users.

\subsubsection{Emotional Dependency and Harm.}
While high levels of emotional engagement can enhance user trust and perceived support, they also create risks of unhealthy dependency. 
Studies of social chatbots demonstrate that emotionally rich interactions may blur the boundaries between human-AI and human-human relationships, fostering attachment patterns that can lead to distress, manipulation, or even severe psychological harm~\cite{zhang2025dark}. 
These dynamics are particularly evident in the case of ``virtual companions,'' where users often attribute human-like needs and emotions to the system, amplifying both the benefits of companionship and the vulnerabilities of over-reliance~\cite{laestadius2024too}.
Mitigation strategies include establishing rigorous ethical guidelines with continuous monitoring, and deploying real-time user reporting mechanisms with escalation protocols to human moderators for high-risk scenarios~\cite{thieme2023designing}.

\subsubsection{Emotion Recognition.}
Accurate emotion recognition represents a critical challenge for CAI4SG applications, where misinterpretation of user emotional states can undermine trust and potentially cause harm in vulnerable populations. 
Current emotion recognition methods face significant limitations in multi-modal integration, typically relying on text-based sentiment analysis while overlooking paralinguistic features such as vocal tone and speech patterns that convey emotional nuance~\cite{kalateh2024systematic}.
Promising mitigation approaches include developing multimodal techniques that integrate text, voice, and contextual information, implementing cultural adaptation mechanisms, and establishing systematic evaluation frameworks across diverse demographic groups to ensure equitable performance in social good applications~\cite{yang2022emotion}.

\subsubsection{Privacy and Security. }
The handling of sensitive user data~\cite{lee2024deepfakes}, particularly in emotionally engaged health applications and roles that invite deep self-disclosure~\cite{lee2020hear}, raises significant privacy and security concerns. 
These concerns extend beyond the technical challenges of data sharing, secure storage, and third-party processing, to encompass broader questions of accountability and user trust. 
Such issues are especially salient for highly engaging roles such as “empathetic supporters” and “virtual companions,” where users often disclose intimate health, relational, and identity-related information~\cite{laestadius2024too, zhang2024ai}. 
Potential solutions include implementing robust encryption protocols, establishing transparent user interfaces that clearly communicate data practices, and developing privacy-preserving technologies to protect user information while maintaining system functionality~\cite{khalid2023privacy}.

\subsubsection{Personalization.}
Personalization is indispensable in emotionally intensive CAI because stable user models can tailor tone, pacing, and content to users’ histories and sensitivities. 
Yet unbounded personalization can overwhelm through excessive tailoring and emotional mirroring, and can drift into sycophancy, where systems are "overly flattering or agreeable" \cite{openai2024sycophancy} to please a user, rather than being accurate~\cite{zhang2025dark}.
In high emotional engagement contexts, these dynamics intersect with known risks of over-reliance and fragile conversational continuity.
Evidence further suggests that sycophancy can degrade reliability and amplify misinformation or discriminatory biases by preferentially echoing a user’s stance.
A practical mitigation is to replace pure preference-following with principle-grounded reward shaping and explicit calibrated disagreement objectives to privilege honesty over agreement and penalize belief-congruent responses in the reward model~\cite{malmqvist2025sycophancy,bai2022constitutional}.

\subsection{Automation, Technical, and Interactional Hurdles}

In high-autonomous CAI roles, automation, technical, and interactional hurdles become particularly pressing because the system assumes greater responsibility for decision-making and service delivery. When autonomy is high, errors in reasoning, system reliability, or user interaction design can propagate without immediate human correction, magnifying risks of misinformation, mis-coordination, or unintended harm in socially critical contexts.

\subsubsection{Algorithmic Bias.}
CAI models are trained on large-scale data that often reflect historical, cultural, and social inequalities, making them prone to inheriting and amplifying biases. 
As demonstrated in studies of rating platforms, even subtle algorithmic design choices can systematically skew outputs, favor certain groups, and mislead users, resulting in inflated scores, distorted perceptions, and ultimately broken trust~\cite{eslami2017careful}.
When CAI systems are deployed in socially critical contexts particularly in high-autonomy roles or across diverse user group, such biases risk exacerbating inequities and overlooking the needs of marginalized communities.
Because fairness is not automatically guaranteed by scale, bias-awareness and mitigation must be built into the design process, for example through algorithmic audits, bias-aware user reporting mechanisms, and ``actionable transparency'' that allows affected communities to recognize and contest biased outcomes~\cite{sakib2024challenging}.

\subsubsection{Explainability and Transparency.}
The ``black box'' character of many CAI systems hinders accountability and erodes public trust, particularly in domains such as mental health support or public service delivery where models may operate with high autonomy and directly affect vulnerable populations~\cite{hepenstal2019algorithmic}. 
Without clarity on how recommendations are generated, users and stakeholders cannot reliably evaluate system reliability or assign responsibility when errors occur.
To address this, CAI should not only provide post-hoc explanations of its outputs but also express uncertainty in ways that are calibrated to its true capabilities~\cite{li2024overconfident,xu2025confronting}. 
Furthermore, empirical studies have shown that interacting with CAI can directly influence users’ own confidence and perceived competence~\cite{li2025confidence}, which underscores the need for more cautious and transparent design choices in order to avoid misleading or overempowering users.

\subsubsection{Misinformation and Deception.}
CAI holds a dual capacity: on one hand, it can be deployed to combat misinformation by disseminating timely, verified information and correcting falsehoods; on the other, the same generative and persuasive abilities can be weaponized to spread false narratives, eroding public trust and amplifying harm~\cite{xu2023combating}.
Such practices challenge the integrity of CAI roles across domains, implicating both autonomy and ethical use. 
Addressing these risks requires embedding verifiability mechanisms, calibrated disclosure of system competence, and clear governance norms that discourage deceptive or manipulative design choices.

\section{Future Directions}

Finally, we suggest future directions and implications in CAI4SG and for the broader AI community that are emerging from current trends, framed by the role-based taxonomy.
In particular, this taxonomy suggests that the central opportunity for CAI4SG is not to improve each role in isolation, but to develop formal mechanisms for determining when a CA should remain within a role versus adapt its autonomy or emotional stance as conditions change. 
This turns the taxonomy into a generative agenda for the AAAI community: it creates concrete openings for work on role boundary specification, role-shift detection, and safety-preserving transitions between roles. 
Such questions map cleanly onto core AI research areas, e.g., formal modelling, uncertainty calibration, multi-agent orchestration, preference learning, and policy specification, and therefore provide a tractable substrate upon which subsequent sections elaborate more domain-specific research directions.

\subsection{Advancing Research in Complex Human-AI Dynamics and Role Evolution}

\subsubsection{Dynamics of Interacting with Multiple CAI Agents and Roles.}
While discrete CA roles provide conceptual clarity, many real-world tasks involve overlapping demands such as simultaneously requiring functional accuracy, empathetic support, and ethical accountability.
In response to this challenge, multiple-agent setups traditionally combine multiple individual agents to increase the reasoning capabilities of AI systems \cite{chen2023reconcile, du2023improving, ge2023openagi} and improve their task performances \cite{hong2024metagpt, qian2023chatdev, xiong2023examining}. However, recent research has investigated the effects of multiple conversational agents on end users. 
% Without requiring strong autonomy or emotional connections with users, 
Increasing the number of agents present in a system can create social influence on users \cite{song2025greater}, causing opinion shifts on matters such as social issues \cite{song2024multi} and artistic choices \cite{song2025more}. This demonstrates the potential of multi-agent systems in the field of persuasive design, in increasing the effectiveness of systems that encourage positive behavior change in areas such as health \cite{balloccu2021unaddressed} or education \cite{ahtinen2020learning}. Another direction of research examines combining multiple CAI agents with different roles. Such systems may reduce cognitive load \cite{jiang2023communitybots}, improving usability and productivity, or improve decision making \cite{park2023choicemates}, leading to better human-AI collaboration outcomes. Major industry players such as Anthropic \cite{anthropic2024claude} and Google \cite{google2024groopy} have responded to this opportunity by launching frameworks that allow developers to easily deploy multi-agent interfaces. 

\subsubsection{Cross-Cultural and Contextualized Research.}

Future work on CAI4SG should expand beyond dominant cultural and demographic contexts to better capture the diversity of user needs and expectations~\cite{liu2024understanding}.
Socio-cultural factors such as communication norms~\cite{qin2025ai}, trust in institutions, or stigma around mental health, can significantly shape how people perceive and interact with conversational agents.
Such contextualized understanding is essential for designing culturally sensitive CAI roles that respect local norms and values, reduce inequities, and ensure that the promise of AI for social good is realized across diverse populations.

\subsection{Strengthening Ethical CAI Frameworks and Governance Across Roles}

\subsubsection{Robust Regulation and Policy.}
As CAI systems expand into socially critical domains, forward-looking regulation is essential to ensure ethical use, fair benefit distribution, and protection of marginalized groups. 
Highly autonomous systems risk amplifying bias and eroding trust if left unchecked, making accountability and transparency requirements central to both public and private deployments~\cite{eslami2017careful}. 
Beyond national guidelines, a multilateral approach is needed: akin to nuclear non-proliferation treaties, advanced AI calls for cross-border governance that establishes shared norms and safeguards against misuse while aligning innovation with the mission of CAI4SG.

\subsubsection{Standardized Ethical Guidelines.}
To ensure that CAI4SG advances equity and trust, there is a pressing need for standardized ethical guidelines tailored to prosocial use cases~\cite{ruane2019conversational}. 
These guidelines should explicitly encompass fairness (avoiding bias and unequal treatment), accountability (clear lines of responsibility when harm occurs), transparency (making system capabilities and limitations visible), and inclusivity (designing with marginalized and diverse communities in mind)~\cite{singhal2024toward}. 
These principles apply across all CAI roles, ensuring that both high-autonomy systems and low-level assistants align with broader societal values and do not inadvertently undermine the very populations they aim to support.

\subsubsection{Data Privacy and Security Enhancements.}
Safeguarding sensitive user information is fundamental to the responsible deployment of CAI4SG. 
Systems must employ robust encryption for both data in transit and at rest, alongside advanced anonymization techniques to minimize the risk of re-identification. 
Transparent user interfaces that clearly communicate what data is collected, how it is processed, and under what conditions it is shared are essential to building trust, particularly in high-stakes domains like healthcare and humanitarian aid. 
Beyond conventional safeguards, adopting federated learning offers a promising pathway for privacy-preserving model training, as it enables insights to be derived without centralizing personal data~\cite{yin2021comprehensive}. 
Together, these measures strengthen both technical resilience and user confidence, ensuring that CAI4SG initiatives protect individuals while delivering equitable social impact.

\section{Conclusion}
Conversational AI for Social Good holds immense transformative potential across domains such as mental health, accessibility, public services, and disaster risk reduction. By leveraging diverse roles differentiated through varying levels of autonomy and emotional engagement, these systems can scale support, expand access, and address urgent societal challenges in ways previously unattainable.

Realizing this potential, however, demands more than technical sophistication. It requires a steadfast commitment to ethical and human-centered design, including transparent and inclusive policies that prioritize human agency and well-being over purely commercial imperatives. Only by embedding fairness, accountability, and inclusivity into every role can CAI achieve sustainable impact without reinforcing existing inequities.

Ultimately, shaping CAI responsibly is not simply a technological challenge but a moral imperative. If guided by robust ethical frameworks and collective governance, CAI can become a catalyst for a more sustainable, just, and healthy future for all.

\section{Acknowledgments}
This work was partially funded by the National University of Singapore CSSH (24-1774-A0002), the National University of Singapore HSS Seed Fund CR (2024 24-1191-A0001), and Google Research Gift.

\bibliography{aaai2026}

\end{document}